\documentclass{article}

\usepackage{latexsym}

\begin{document}

\centerline{\Large{Distribution of amplitude in an extended phase space}}

\centerline{\Large{of a generic quantum system}}

\vskip 0.5 truecm

\centerline{C. Lopez}

\vskip 0.3 truecm

 \centerline{\small{Dept. of Physics and Mathematics, Facultad de Ciencias}}
\centerline{ \small{UAH, Alcal\'a de Henares  E-28871 (Madrid, SPAIN)}}
\centerline{ \small{carlos.lopez@uah.es }}

\vskip 0.3 truecm

\begin{abstract}


The action reaction principle is violated by the projection of state in some simple quantum measurements. 
A formulation of Quantum Mechanics in an extended phase space is proposed in order to restore the action reaction principle.
All predictions of the standard theory are reproduced. Observable effects of an accompanying
de Broglie wave are also predicted.


\vskip 0.3 truecm

\end{abstract}	

\vskip 0.5 truecm

\section{Introduction}

The action reaction principle (ARP) is as firmly established as the conservation of energy.
Theoretical models where one or the other is not fulfilled have most pro\-ba\-bly incomplete
phase spaces,  and additional  variables in an extended phase space will restore them. The axiom of measurement 
in Quantum Mechanics (QM) violates the ARP in  simple examples. This paradox 
would not appear if the phase space of standard QM were incomplete. The 
wave function collapse, projection of state, generates many paradoxes in the interpretation of QM,
and a possible incompleteness of the theory was already considered in the seminal paper \cite{EPR}.

Alternative formulations and interpretations  incessantly appear although the scientific success of QM is overwhelming;
see e.g.  \cite{biblioguide} for a detailed bibliographic list up to 2004. No--go mathematical results \cite{Bell}  \cite {GHSZ}
rule out alternative formulations with hidden variables fulfilling the hypothesis of the theorems. On the other hand,
alternatives as Bohmian mechanics \cite{Bohm} are explicit examples of consistent models with hidden variables, 
obviously not fulfilling the hypothesis of those no--go theorems.

With regards to Bell's theorems, there has been a wide research activity around which of the hypothesis is (or are) not fulfilled by the laws of Physics. Logic, axioms of mathematical probability, even realism, have each open more or less fruitful lines of study in different fields, although none seems to be relevant in fundamental Physics. 

Non locality is acknowledged as the 
most probable property of the real world that explains the violation of Bell's inequalities
in experiments \cite{Aspect}. But that seems to be incompatible with relativity.
The non local interaction between measurements of an entangled pair is very different from 
known fundamental interactions (electromagnetic, gravitational, \ldots). It acts between events either spatially or causally separated, it does not decay with distance and it is specifically directed towards the entangled pair. No theoretical model of its mechanisms (as, e.g., gauge theories) has been proposed, neither an isolated effect of an hypothetical signal over an apparatus has ever been detected.

Standard Quantum Mechanics is non local through the
projection of state, the same axiom that crashes with the ARP,
unitary evolution, etc. It seems reasonable to look at the established theory in order to find some other ingredient of the mathematical formalism that could be the cause of violation of Bell's inequalities, instead of a non local projection of state.

Born's rule,  a distribution of probability obtained from a distribution of amplitude, is not considered in the  hypothesis of Bell's theorems; e.g., typical interference phenomena as in the two slit experiment can not be reproduced with a distribution of probability 
\footnote{In a minimum of the diffraction pattern there is no way to get a near vanishing total probability by addition of two positive independent ones, which is a kind of trivial Bell's type inequality.}.
A window is open to assign the violation of Bell's inequalities in Nature to the description of physical states through distributions of amplitude, instead of non locality of the wave function collapse.

Violation of the ARP in the projection of state suggest to extend the phase space. An extended Hilbert space will necessarily contain
elementary states with joint precise values of non commuting magnitudes.
It seems that such states would violate the uncertainty principle. The point is if the uncertainty principle applies to the description of states or it just determines a fundamental limit of accuracy to the knowledge (through measurement) of conjugate variables.

Commutation relations apply to
evolution in Hamiltonian dynamics. In Classical Mechanics, functions on the phase space (e.g., conjugate variables) commute under the usual product, and do not generically commute under the Poisson bracket operation \footnote{This fact necessarily limits accuracy of classical joint measurements, although there is  no Plank's constant as fundamental lower bound.}. The existence of quantum
states with joint precise (hidden) values of non commuting magnitudes could be consistent  with an uncertainty principle for joint measurement of these magnitudes.

In \cite{mio}, I presented a mathematical framework with an extended  phase space for spin variables,
consistent with the standard theory. The model is inspired in the 
 path integral formalism, where I consider the formal hypothesis that
there are families of virtual paths characterized by joint precise final values of non 
commuting physical magnitudes, as position and momentum $(x,p)$, or spin in two or more directions $(s_1,s_2, \ldots )$. 
Sum of  $exp(i S/\hbar )$, $S$ the path integral, for all paths in the family determines
a distribution of amplitude $\Psi (x,p)$ or 
$\Psi (s_1,s_2, \ldots )$ in an extended phase space.
The orthodox state in the standard phase space is then obtained by addition of amplitudes,
i.e.,  computation of marginal amplitudes, for all
fa\-mi\-lies with common value of the considered magnitude, e.g., $\Psi (x) = \int dp\,\, \Psi (x,p)$.
Bell's type inequalities do not apply; the fundamental ingredient is,
as in the usual formulation,  a distribution of amplitude (here  in an extended phase space with hidden variables),
and not a distribution of probability,

The main physical hypothesis of the proposed theory, introduced to restate the ARP, are the existence of an accompanying 
subsystem, the de Broglie wave \cite{deBroglie}, and  additional variables for the corpuscular (sub)system.
Elementary states of a quantum system are characterized by a family of precise joint values 
of all physical magnitudes representing
the particle (corpuscle) behavior, and a distribution of amplitude (ray in a Hilbert space)
representing its wave like behavior. Bell's experiments 
with spin variables on a sample of entangled pairs \cite{Bell} (and Aspect's for polarization variables \cite{Aspect}) 
appear in this 
theory as local interference phenomena, analogous to the two slit experiment.

The aim of this paper is to apply the former hypothesis to a generic quantum system and arbitrary family of magnitudes.
Section $2$ reviews the physical arguments, the need to extend the phase space in order to preserve the ARP.
In Section $3$ the mathematical framework of the theory is presented,  and its correspondence with  standard QM 
established. 
Section $4$ develops the description of composite systems and entanglement. Interference terms in the marginal amplitudes give account of the 
apparently non local correlation between measurements.
Particular cases, the two slit experiment and the phase space of spin variables, are presented in Section $5$.


\section{Quantum measurement and the action reaction principle}

The ARP represents, in its most generic terms,  a basic hypothesis of dynamical theo\-ries.
An isolated composite system evolves, with regards to some coarse--grained 
variables, as a free elementary system. Internal variables must  be correlated: 
departure from the free dynamics of a subsystem 
is accompanied by a corresponding departure of the other. Third Newton's law
is necessary for the consistency of Classical Mechanics; otherwise, the first law 
would not have been established, because  elementary classical systems
are composite at a microscopic level.  The ARP is independent of the particular type 
of interaction  between subsystems; it belongs to the foundations of Physics and its generic framework
(ideal isolated systems, free dynamics, \ldots). 

Friction forces apparently violate the conservation of energy; obviously, there are additional 
(microscopic) variables giving account of the lost energy in the balance, and all known fundamental interactions are 
conservative. In Brownian motion,  the pollen grain receives impulses 
from  unobservable systems, the fluid molecules; once incorporated the corresponding variables in an extended phase space,
an opposite impulse balances the momentum equation.

The ARP and the axiom of quantum measurement are contradictory. One or the other must be rejected
in its present formulation. Let $|a_1>$ be the state of a quantum system, eigenstate of a physical magnitude $A$.
When $A$ is measured, the result of measurement is $a_1$ with certainty, and the final state of the system is 
again $|a_1>$, with trivial projection of state. The pointer of the measurement apparatus has changed of state, from a neutral position to the result
``$a_1$''. ARP is violated.

In a theory with additional variables in an extended phase space, elementary states of the quantum system should be
described by joint precise values of non commuting magnitudes, because maximal families of compatible magnitudes are already
considered in the standard formulation. In the orthodox  Hilbert space $\cal H$
there are not common eigenstates for incompatible magnitudes; an extended Hilbert space ${\cal H}_{ext}$, with different representation of physical magnitudes, will be the first ingredient of the proposed framework.

Let $c_1|a_1>$ $+$ $c_2|a_2>$ be the initial  state of a quantum system, an elementary particle, 
and suppose wave packets of the two components are spatially separated 
(e.g., with  a Stern Gerlach apparatus). A particle detector is located
in one of the ``virtual'' paths, $|a_1>$, and the result of measurement is negative 
\footnote{It will happen with relative frequency 
$|c_2|^2$.}. The new, projected state of the system is $|a_2>$. The detector has not, apparently, changed of state, and the ARP is violated in this indirect measurement.

Obviously, the detector is designed to show an observable response when interacting 
with a particle located in the surrounding spatial region.
Perhaps some type of non local interaction between particle at $|a_2>$ region and detector at $|a_1>$ happens,
an interaction that does not generate an observable response in the detector.

I will consider an alternative hypothesis: the detector interacts with a system spatially located in its neighborhood.
This hypothetical system is not the particle, which is certainly located away. An hypothetical wave like system, which will be denoted the de Broglie wave, does not generate the same reaction than a corpuscular system, and another type of detector should be designed to observe a response. An accompanying wave of an isolated particle is a wave in vacuum, the vacuum
becoming a relevant physical ingredient in non relativistic QM, as it is relevant in 
the opposite length scales of Quantum Field Theory (e.g., Casimir energy) and Cosmology (dark energy).

Let us suppose that an elementary particle, and in general a quantum system,
is a composite of a corpuscular subsystem and a wave like subsystem. The corres\-ponding phase space 
of the composite system must describe states of both subsystems. If the standard representation
$c_1|a_1>$ $+$ $c_2|a_2>$ in $\cal H$, or the co\-rres\-pon\-ding vector  in an 
extended Hilbert space ${\cal H}_{ext}$, is associated to the de Broglie wave 
(as seems to suggest interaction of the detector with the  $|a_1>$ wave component and not 
with the particle), 
additional variables of state for the corpuscular component
(known to be located at the spatial region of the other
wave packet $|a_2>$) must be considered
for a total phase space ${\cal H}_{ext}^{wave} \times {\cal P}^{corp}$.


\section{Extended phase space}

We must confront the standard phase space of a quantum system, the Hilbert space 
$\cal H$, with an extended phase space ${\cal H}_{ext}^{wave} \times {\cal P}^{corp}$
describing both components, wave like and corpuscular subsystems. First, a formal path integral formalism
is considered. Then, the definition of an extended phase space for a generic quantum system 
$\cal S$ and 
arbitrary family of physical magnitudes $\cal F$ is given. A correspondence between states in both formalism
determines equivalence of the theories, with regards to phase space description.

\subsection{Abstract path integral formalism}

$\cal S$ denotes a quantum system,  ${\cal F} = \{ A, B, C, \ldots \}$ a generic family of $N$ physical magnitudes (observables) of
$\cal S$, $M_A = \{ a_i \}$ the set of possible values of magnitude $A$,
$M_B = \{ b_j \}$ of $B$, etc. Additional magnitudes can be incorporated if needed; in many 
cases some magnitudes are ignored, when they are not correlated to those of our interest \footnote{Magnitudes in $\cal F$ are not necessarily functionally independent, as would be a  coordinate  description of classical phase space. Angular momentum, for example, is a vector
but components in three independent directions
 do not exhaust the quantum information, because  operators in additional directions do not commute with the selected three.}. 
An element $\lambda \in M_{\cal F} \equiv $ $M_A \times M_B \times M_C \cdots $, 
$\lambda = (a_i,b_j,c_k, \ldots )$, is a $N$--tuple of  values of all magnitudes in 
$\cal F$.

Let us consider an abstract, formal set of virtual paths, such that
we can assign to each of them an action integral $S(path)$ and a final value 
$\lambda$ \footnote{A specific description of the set of paths and rules of assignment should be given if no alternative way to determine states is available.}. $[path](\lambda )$ denotes the subset of paths with final value $\lambda = (a_i,b_j, \ldots )$. We formally define the amplitude

\begin{equation}
Z(\lambda ) = \sum _{[path](\lambda )} e ^{\frac {i}{\hbar}S(path)}
\end{equation}

It is possible to fix the values  of a subset of magnitudes 
${\cal F}_1 \subset {\cal F}$,
and define $Z(\lambda _1 )$, $\lambda _1 \in M_{{\cal F}_1}$, by addition of elementary terms
$exp(i S/\hbar )$ for all paths with common values $\lambda _1$.
The amplitude $Z$ associated to a family of paths union of disjoint subfamilies with corresponding amplitudes $Z_1$, $Z_2$, \ldots,
is  $Z= Z_1 + Z_2 + \cdots$. For example,
$Z(\lambda _1 ) = \sum _{\pi _1 (\lambda ) = \lambda _1} Z(\lambda )$ in the previous case 
$\lambda _1 \in M_{{\cal F}_1}$,
with $\pi _1$ the natural projection  $\pi _1 : M_ {\cal F} \to M_{{\cal F}_1}$.

In the standard 
treatment \cite{Feynman}, the family ${\cal F}_1$ considered is one made of compatible magnitudes, with vanishing commutation relations, e.g. final position coor\-dinates.
A quantum state  of $\cal S$ is completely determined through a distribution of amplitude $Z(\lambda _1)$,
which defines a vector $|S> = \sum _{\lambda _1} Z(\lambda _1) |\lambda _1>$ in the co\-rres\-pon\-ding Hilbert space $\cal H$,
generated by elementary states $|\lambda _1> = |(a_i,b_j,\ldots )>$, common 
eigenvectors of the family of self adjoint operators representing the magnitudes in ${\cal F}_1$.

Another family of  magnitudes ${\cal F}_2$ (compatible among themselves, but not with all of 
${\cal F}_1$) 
has its corresponding basis of eigenvectors, and
the relations between bases of eigenvectors $|\lambda _1>$ and $|\lambda _2>$ is consistent with the commutation relations 
(quantization rules) of operators. Given the change of bases the new distribution $Z(\lambda _2)$ is obtained from 
$Z(\lambda _1)$,
$Z(\lambda _2) =$ $\sum _{\lambda _1}$ $<\lambda _2|\lambda _1>$ $Z(\lambda _1)$.

In the abstract path integral formalism, we can express the  amplitude $Z(\lambda _2)$
as 

\begin{equation}
Z(\lambda _2) = \sum _{\lambda _1} 
\sum _{[path](\lambda _1)\cap [path](\lambda _2)} 
e ^{\frac {i}{\hbar}S(path)}
\end{equation}
where the set of paths $[path](\lambda _2)$ has been decomposed into disjoint subsets
$[path]$ $(\lambda _1)$ $\cap$ $[path](\lambda _2)$.
Denoting $Z(\lambda _1,$ $\lambda _2)$ the  amplitude associated to the family of paths 
$[path](\lambda _1)$ $\cap$ $[path](\lambda _2)$, 

\begin{equation}
Z(\lambda _1,\lambda _2) = \sum _{[path](\lambda _1)\cap [path](\lambda _2)} 
e ^{\frac {i}{\hbar}S(path)}
\end{equation}
we can obtain both $Z(\lambda _1)$ and $Z(\lambda _2)$ as marginal amplitudes

\begin{equation}
Z(\lambda _1) = \sum _{\lambda _2} Z(\lambda _1, \lambda _2)
\quad 
Z(\lambda _2) = \sum _{\lambda _1} Z(\lambda _1, \lambda _2)
\end{equation}
In this way, we can get any orthodox representation of a state (associated to a family of compatible magnitudes)
from a 
representation in an extended phase space (associated to a larger family of non commuting magnitudes)
\footnote{Families ${\cal F}_1$ and ${\cal F}_2$ can contain 
a common subfamily ${\cal F}_c = {\cal F}_1 \cap {\cal F}_2$ of magnitudes;
the set of paths characterized by joint parameters $\lambda _1$ and 
$\lambda _2$ which do not share the same values in ${\cal F}_c$ is void. Both 
$\lambda _1$ and $\lambda _2$ must project onto a common $\lambda _c$.}. 

We consider next the formulation of a non relativistic Quantum Mechanics 
inspired in the former abstract path integral formalism,  making the hypothesis that a
distribution of amplitude defining a vector (ray) in an extended Hilbert space
represents a state of the quantum system.
Being inspired in the path integral formalism,  a relativistic quantum theory
in an extended phase space
could be formulated when a Lorentz invariant action were considered. Notice that
virtual  paths arriving to a space time event are contained inside its past 
light cone, i.e.,  acausal interactions do not appear through the amplitudes
obtained in this way.

\subsection{Extended phase space}

If the phase space of the quantum system is to be extended, the family of magnitudes considered 
in $\cal F$ must contain non commuting operators. In the former abstract path integral, a distribution of amplitude $Z(\lambda )$ 
determines the  quantum state, where $\lambda \in M_{\cal F} $ is a $N$--tuple of 
values of magnitudes in $\cal F$.   

We define a Hilbert space ${\cal H}_{ext}$, generated by  vectors $\|\lambda> =$ 
$\|(a_i,b_j,$ $c_k,$ $\ldots )>$,
orthonormal  vectors representing elementary states of $\cal S$ with joint precise values of all magnitudes in $\cal F$. 
Operators representing the magnitudes (denoted with the same symbols) $A$, $B$, etc.,
have the basis $\|\lambda >$ of common eigenvectors, with eigenvalues $a_i$, $b_j$, \ldots 
$\|S> = \sum _{\lambda } Z(\lambda ) \|\lambda >$ is a generic vector of state in 
${\cal H}_{ext}$.

By construction, $A$, $B$, etc., commute. This  represents the fact that elementary states of the system have joint precise values of all considered magnitudes.
Together with operators $A$, $B$, etc. introduced in the 
definition of the phase space ${\cal H}_{ext}$, we will need additional  operators $A^d$, 
$B^d$, \ldots, acting on  ${\cal H}_{ext}$ to represent the 
dynamical role of  magnitudes, e.g., when they appear in the Hamiltonian. The ``phase space'' 
representation  $A$, $B$, etc., preserves
the commutation property of usual multiplication of functions, while 
``dynamical'' representation  
$A^d$,  $B^d$, \ldots,
would fulfill the usual quantization rules. Dynamics is not considered in this article, 
only an alternative phase space representation of quantum states is analyzed.

Together with the extended Hilbert space ${\cal H}_{ext}$, we introduce a label 
$\lambda _0 \in  M_{\cal F}$ characterizing the physical state of the corpuscular component of $\cal S$.
Both subsystems, de Broglie wave and particle, are  described by  $(\|S>, \lambda _0)$
in  a complete phase space ${\cal H}_{ext} \times M_{\cal F}$. 
$\lambda _0$ determines the value of an arbitrary measurement on  state $(\|S>, \lambda _0)$; 
the result of an $A$ measurement is $\pi _A (\lambda _0)$ (some $a_i$), where 
$\pi _A :  M_{\cal F} \to M_A$ is the natural projection.

$\lambda _0$ is  hidden, we can at most know precise values for a maximal family of compatible magnitudes, 
a subfamily of $\cal F$. The observable relative frequencies for an arbitrary measurement on an ensemble $\|S>$
of states $(\|S>, \lambda _0)$, i.e., with common component $\|S>$ and all allowed values of 
$\lambda _0$,
is obtained from  the distribution of amplitude through Born's rule. 

Born's rule \cite{Born} at ${\cal H}_{ext}$ is defined in two steps  as follows. If we want to obtain a (perhaps formal)
distribution of probability $P(\lambda _1)$ for a subfamily ${\cal F}_1 \subset {\cal F}$, 
 ${\cal F}_1 =$ $\{ D,E,G, \ldots \}$, 
$\lambda _1 \in M_{{\cal F}_1}$, 
we first project $\|S>$ onto the Hilbert space generated by vectors $\|\lambda _1>$,
$\tau _1 : {\cal H}_{ext} \to {\cal H}_1$, 
$\tau _1(\|\lambda>)$ $=$ $\|\pi _1(\lambda )>$ $=$ $\|\lambda _1>$, i.e.,

\begin{equation}
 \tau _1(\|(a_i,b_j,c_k,d_l, e_m, g_n, \ldots)>) 
= \|(d_l, e_m, g_n, \ldots) >
\end{equation}
with $\lambda = (a_i,b_j,c_k,d_l, e_m, g_n, \ldots) \in  M_{\cal F}$ and $\lambda _1 =$ 
$(d_l, e_m, g_n, \ldots)  \in M_{{\cal F}_1}$.

For $\|S> = \sum _{\lambda } Z(\lambda ) \|\lambda >$, the  projection $\|S_1>$ is

\begin{equation}
\|S_1> = \tau _1 (\|S>) = \sum _{\lambda _1} \left(   \sum _{\pi _1(\lambda ) = \lambda _1} Z(\lambda )
\right) \|\lambda _1>
\end{equation}
The coefficient $Z(\lambda _1)$ of an elementary state
$\|\lambda _1>$ in ${\cal H}_1$ is the marginal amplitude obtained from $Z(\lambda )$,

\begin{equation}
Z(\lambda _1) = \sum _{\pi _1(\lambda ) = \lambda _1} Z(\lambda )
\end{equation}
In the abstract path integral, this is the union of disjoint families of virtual paths
characterized by different $\lambda$ in a larger family characterized by a common $\lambda _1$.

The second step is the standard  $P(\lambda _1) = |Z(\lambda _1)|^2$. Obviously,
in general $|Z(\lambda _1)|^2 \neq$ $\sum _{\pi _1(\lambda ) = \lambda _1}$ 
$|Z(\lambda )|^2$, that is,
the probability distribution $P(\lambda _1)$ does not match the
marginal probability distribution $P'(\lambda _1) =$ $\sum _{\pi _1(\lambda ) = \lambda _1}$ $P(\lambda )$
obtained from the formal probability distribution $P(\lambda )$ $=|Z(\lambda )|^2$ associated to the amplitude distribution $Z(\lambda )$. 

The former Born's rule can be interpreted as
a representation of the correlation (or interaction) between the corpuscular and wave like subsystems. Contextuality of QM
is in this formulation a consequence of interference, with different inter\-fe\-rence 
results for different marginal amplitudes. 
The interference that appears in  marginal amplitudes  can not be generically reproduced with marginals of a distribution of probability in $M_{\cal F}$, as Bell's  theorems show.

The formal $P(\lambda)$ is not observable, no joint measurements of all magnitudes can be consistently performed. The unobservable $P(\lambda )$ does not reproduce observable probabilities through marginal probabilities. For example, in the two slit experiment the diffraction pattern is not obtained from a sum of probability distributions for each individual slit; interference, or sum of wave like degrees of freedom, is a more adequate analogy. Generically, an observable $P(\lambda _1)$ distribution is calculated through the sum of wave like degrees of freedom in the marginal amplitude and Born's rule. According to the path integral point of view, all amplitudes are calculated in a strictly causal way, through
integral along virtual paths inside the past light cone.

\subsection{Correspondence}

For a quantum system $\cal S$, let ${\cal F}$ be a family of  physical magnitudes,
and ${\cal H}_{ext}$ the associated extended Hilbert space, generated by the basis
$\{ \|\lambda > \}$. ${\cal H}_{QM}$ will denote the standard Hilbert space, 
orthodox quantum phase space of $\cal S$. For each 
orthonormal basis $\{ |\lambda _1> \}$ in ${\cal H}_{QM}$ of common eigenvectors of a 
maximal family of compatible operators  in a subfamily ${\cal F}_1 \subset {\cal F }$, we  describe next the mathematical conditions for
a co\-rres\-pon\-dence between vectors $\|S> \in {\cal H}_{ext}$ and 
$|S> \in {\cal H}_{QM}$ representing the same physical state (or ensemble) of $\cal S$.

The most direct (and strongest) conditions are simply the equations
($\pi _1(\lambda ) = \lambda _1$)

\begin{equation}
\tau _1 (\|\lambda >) \equiv \|\lambda _1> = |\lambda _1>
\quad
\tau _1 (\|S>) = |S>
\end{equation}
Marginal amplitudes of $Z(\lambda )$ will match $z(\lambda _1)$,
for $\|S> = \sum _{\lambda } Z(\lambda ) \|\lambda >$,
$|S> = \sum _{\lambda _1} z(\lambda _1) |\lambda _1>$, i.e.,
$\sum _{\pi _1(\lambda ) = \lambda _1} Z(\lambda ) = z(\lambda _1)$.

However, the observable properties of the physical state are determined by the ray
$[|S>] = \{ c |S> \}$ ($c$ arbitrary complex numbers), so that it is enough that rays 
$[\tau _1(\|S>)]$ and $[|S>]$
coincide. The corresponding equations will be projective ones, which I will denote 
with the symbol $::$. Maintaining the identification $\|\lambda _1> = |\lambda _1>$
we have

\begin{equation}
\sum _{\pi _1(\lambda ) = \lambda _1} Z(\lambda ) :: z(\lambda _1)
\end{equation}

The mildest way of the correspondence takes into account the freedom in phase
that exists in the definition of a basis of  eigenvectors,
even after normalisation. We could state, for each subfamily ${\cal F}_1$
and each one dimensional eigenspace $[|\lambda _1>]$,
a relation 

\begin{equation}
\|\lambda _1 > = e^ {i\theta (\lambda _1)}|\lambda _1>
\end{equation}
which determines the projective equation

\begin{equation}
|\sum _{\pi _1(\lambda ) = \lambda _1} Z(\lambda )| :: |z(\lambda _1)|
\end{equation}
These equations should be solved for the $Z(\lambda)$ (defining $\|S>$),
given the  $z(\lambda _1)$, $z(\lambda _2)$, $z(\lambda _3)$, \ldots of $|S>$ for all 
maximal subfamilies ${\cal F}_1$, ${\cal F}_2$,${\cal F}_3$, \ldots  $\subset {\cal F}$ of compatible magnitudes \footnote{Equivalently, given 
$z(\lambda _1)$ and the
change of bases of eigenvectors in ${\cal H}_{QM}$ 
for all maximal compatible subfamilies ${\cal F}_2$, ${\cal F}_3$, \ldots  of  the considered physical magnitudes ${\cal F}$.}. 
The abstract path integral suggest there could be solution in a generic physical
case; I do not have proof of that. Equations are linear (projective),
and obviously no positivity condition appears, amplitudes are not even real numbers
\footnote{Wigner's quasiprobability distribution \cite{Wigner} in the phase space of a point particle is an example of partial solution for the analogous problem of existence of a distribution of probability in an extended phase space, with hidden variables. The positivity requirement for probabilities can not be generically fulfilled.}.

A correspondence for the dynamical operators $A^d$ would also be needed.
In particular, an extended Hamiltonian in ${\cal H}_{ext}$ should
determine a dynamics $\|S>(t)$ compatible with the standard $|S>(t)$ 
through the previous correspondence of states.

There is not a correspondence for the corpuscular variables 
$\lambda _0$, which do not exist in the orthodox theory.
We could incorporate hidden $\pi _1(\lambda _0)$ labels (for each
set ${\cal F}_1$ of compatible magnitudes) to the interpretation of the
standard theory,  without observable consequences. 

An interesting property of this formulation  is
the classical limit. Classical limit for an elementary particle 
has no physical relevance, Plank's constant can not be driven to $0$,
and quantum effects from the wave component are unavoidably relevant. The relevant classical limit applies to complex systems, with many components. For them,
only coarse--grained, global, macroscopic variables are observable. Classical
additive  state variables of the corpuscular component survive in the classical limit, while 
wave variables become irrelevant and we can apply $\hbar \to 0$. In particular, only corpuscular variables are observable in the interaction with a macroscopic system as a measurement apparatus.

The quantum Hamiltonian $H_0$ for $\lambda _0(t)$ evolution can be 
rewritten as $H_0 = H_{class} + V_{QM}$, with $H_{class}$ the classical Hamiltonian.
$V_{QM}$ is  the (possibly stochastic) interaction term (quantum potential in Bohm's formulation \cite{Bohm})
between corpuscular and wave subsystems, 
and also disappears in the classical limit. 

Quantum interaction between particle and vacuum wave is out of reach of observation, we can not 
analyse its deterministic or genuinely probabilistic character. Therefore, the observable evolution must be described in a probabilistic formulation, that incorporates wave like 
effects through amplitudes and Born's rule. In the classical limit, once these effects become negligible for 
global state variables, we get a deterministic theory.

An observable
ingredient that does not exist in the orthodox theory is the de Broglie wave subsystem;
recall it was introduced in order to preserve the ARP in indirect measurements, when particle and 
detector in different spatial regions can not interact, if we maintain the hypothesis of locality.


\section{Composite system}

Let $\cal S$ be a composite quantum system, with components (subsystems) ${\cal S}_I$ and
${\cal S}_{II}$, $\cal F$  a family of physical magnitudes of $\cal S$,
${\cal F}_I$ magnitudes specific of subsystem ${\cal S}_I$, ${\cal F}_{II}$ of 
${\cal S}_{II}$,
and ${\cal F}_{int}$ magnitudes, as a potential of interaction, defined on the composite.
As before, a Hilbert space ${\cal H}_{ext}$ is defined through
an orthonormal  basis of vectors $\|\lambda> =$ $\|\lambda _I>$ $\|\lambda _{II}>$
$\|\lambda _{int}>$, $\lambda _I \in M_{{\cal F}_I}$, etc.

For example, two independent systems with vectors of state

\begin{equation}
\|S_I> = \sum _{\lambda _I} Z_I(\lambda _I) \|\lambda _I>
\quad 
\|S_{II}> = \sum _{\lambda _{II}} Z_{II}(\lambda _{II}) \|\lambda _{II}>
\end{equation}
define a state of the composite

\begin{equation}
\|S> = \sum _{\lambda = (\lambda _I, \lambda _{II}}
Z_I(\lambda _I) Z_{II}(\lambda _{II}) \|\lambda _I>  \|\lambda _{II}>
\end{equation}

In general, when both subsystems interact the state of the composite will not be a direct product, but some

\begin{equation}
\|S> = \sum _{\lambda } Z(\lambda ) \|\lambda>
\end{equation}
in which additional magnitudes of interaction could be taken into account, and appear in 
$\lambda$. 

If $A_I$ and $B_{II}$ are a maximal family of compatible operators in ${\cal F}_I$ and ${\cal F}_{II}$ respectively
(magnitudes of a system trivially commute with magnitudes of the other), the projection 
of $\|S>$ onto ${\cal H}_{A_I} \times {\cal H}_{B_{II}}$ is

\begin{eqnarray}
\tau _{A_IB_{II}}(\|S>) =
\sum _{i,j} \left(
\sum _{\pi _{A_I}(\lambda ) = a_{Ii}, \pi _{B_{II}}(\lambda ) = b_{IIj}}
\right)
\|a_{Ii}>\|b_{IIj}> = \\
\sum _{i,j} Z(a_{Ii},b_{IIj}) \|a_{Ii}>\|b_{IIj}>
\end{eqnarray}
with $Z(a_{Ii},b_{IIj})$ the corresponding marginal amplitudes. 

Correspondence with the standard formalism will be fulfilled if the previous state
is an equivalent representation of the orthodox $|S> = \sum _{i,j}$ $c_{ij}$ $|a_{Ii}>$ 
$|b_{IIj}>$.
A state is said to be entangled if $c_{ij}$ does not factorize, or in the 
extended formalism, if $Z(\lambda )$ does not factorize. 

Relative frequencies are obtained as usual, $P(a_{Ii},b_{IIj}) = |c_{ij}|^2$ in the standard 
formulation when $|S>$ is normalized, and 
$P(a_{Ii},b_{IIj})$ $::$ $|Z(a_{Ii},b_{IIj})|^2$ in the alternative formulation.
Obviously, $P(a_{Ii})$ is obtained as a marginal probability
$P(a_{Ii}) = \sum _j P(a_{Ii},b_{IIj})$, and it is independent of the
chosen magnitude in the second system, e.g.
$P(a_{Ii}) = \sum _k P(a_{Ii},c_{IIk})$ for a magnitude (more precisely, maximal family)
$C_{II}$.

The (extended) state of system $\cal S$ is determined by some vector $\|S>$ and 
 label $\lambda _0 = $ $(\lambda _{I0}, \lambda _{II0})$, if we ignore for simplicity
magnitudes ${\cal F}_{int}$, for example because after interaction both subsystems are far apart. $\lambda _0$ determines the result of an arbitrary measurement, or pair of independent measurements on each subsystem. When measuring system ${\cal S}_I$ its
new label $\lambda  _{I0}'$ will change (magnitudes non commuting  with the one measured will
evolve along the measurement interaction), but this does not modify
the label $\lambda _{II0}$, because all magnitudes in ${\cal F}_{II}$ commute with the
measured magnitude in ${\cal F}_I$. With regards to the
relative frequencies, as we said $P(b_{IIj})$ is independent of 
magnitude $A_I$ measured, although there will be in general a correlation
in $P(a_{Ii},b_{IIj})$. 

If we understand the amplitudes $Z(\lambda )$ as  result of a relativistic path integral calculation, 
virtual paths characterized by $\lambda = (\lambda _I, \lambda _{II})$
are contained in the past light cones of events corresponding to $\lambda _I$
and $\lambda _{II}$, and in the shared past region 
of both there will be  contributions  of interaction terms
in the
action integral. This is the origin of correlation. After both subsystems
separate, do not interact, the additional contributions along each individual path 
will  not modify the established correlation, although obviously they can 
generate individual evolution of each subsystem. That is, all correlation information in $Z(\lambda)$ has a causal origin in the path integral formalism.

Given a magnitude $A_I$ of ${\cal S}_I$ and two non commuting magnitudes
$B_{II}$ and $C_{II}$ of ${\cal S}_{II}$ we can calculate
the marginal $Z(a_{Ii},$ $b_{IIj},$ $c_{IIk})$, and a formal 
$P(a_{Ii},$ $b_{IIj},$ $c_{IIk})$. Again, the marginal $P'(a_{Ii},b_{IIj})$
$= \sum _k P(a_{Ii},b_{IIj},c_{IIk})$ will not match generically
the observable $P(a_{Ii},b_{IIj}$, because of interference in the marginal amplitude
$Z(a_{Ii},b_{IIj}) =$ $\sum _k$ $Z(a_{Ii},b_{IIj},c_{IIk})$.

Let us consider a correlated magnitude $C_T = C_I + C_{II}$, with value $c_T$, as a consequence of a
past interaction. The label $\lambda _0$ (corpuscular degrees of freedom) will fulfil the equation

\begin{equation}
\pi _{C_T}(\lambda _0) = c_T = \pi _{C_I}(\lambda _0) +
\pi _{C_{II}}(\lambda _0) = c_{Ik} + c_{IIk}
\end{equation}
so that perfect correlation appears in a measurement of 
magnitudes $C_I$ and $C_{II}$ on the entangled pair. 
A subset $M_{corr} \subset M_{\cal F}$ determines the correlated values.

Similarly, the vector of state (wave degrees of free\-dom) fulfils
$C_T\|S>$ $=$ $c_T\|S>$. If $\{C_I,$ $C_{II}\}$ is a maximal family of 
compatible magnitudes, the projected state

\begin{equation}
\tau _{C_IC_{II}}(\|S>) =
\sum _{k,l} \left(\sum _{\pi _{C_I}(\lambda ) = c_{Ik},
\pi _{C_{II}}(\lambda ) = c_{IIl}}
Z(\lambda )\right)
\|c_{Ik}>\|c_{IIl}>
\end{equation}
defines, as usual, the marginal amplitudes $Z(c_{Ik},c_{IIl})$, which vanish of 
$c_{Ik} + c_{IIl} \neq c_T$. 

However,
not all terms $Z(\lambda )$ with $\lambda \in M_{\cal F}/M_{corr}$ necessarily vanish, it is enough that 
interference in the marginal $Z(c_{Ik},c_{IIl})$ is destructive. That is, $\lambda _0 \in 
M_{corr}$, but parameter $\lambda$ in the sum defining $\|S>$ belongs to 
$M_{\cal F}$. 
As an example, in the two slit experiment  wave amplitudes 
coming from both slits do not vanish in a zero of the diffraction pattern, 
and a destructive interference of both determines the null probability density there.
A correspondence with the standard state $|S> = 
\sum _k z_k |c_{Ik}>|c_{IIk}>$ ($c_{Ik} + c_{IIk} = c_T$) can now be established,
$|z_k| :: |Z(c_{Ik},c_{IIk})|$.

With another correlated magnitude $D_T = D_I + D_{II}$, with value $d_T$, we can 
define the marginals

\begin{equation}
Z(c_{Ik},c_{IIl},d_{Im},d_{IIn}) \equiv Z_{klmn}
\end{equation}
as usual, and formal, unobservable, probability distributions $P(c_{Ik},c_{IIl},d_{Im},$ $d_{IIn})$ whose marginal probabilities do not match those obtained through Born's rule.

As pointed out before, $Z_{klmn}$ does not vanish generically
for $c_{Ik} + c_{IIl} \neq c_T$ or $d_{Im} + d_{IIm} \neq d_T$. Only the marginals, used in Born's rule to obtain observable re\-la\-tive frequencies, 
$Z(c_{Ik},c_{IIl})$ $=Z_{kl} =$ $\sum _{mn} Z_{klmn} $ will vanish
for $c_{Ik} + c_{IIl} \neq c_T$. Therefore, in a marginal 
$Z(c_{Ik},d_{IIn}) =$ $\sum _{lm}Z_{klmn}$ there will be contributions of
terms $Z_{klmn}$ where $c_{Ik} + c_{IIl} \neq c_T$. We can compare this generic
case with the two slit experiment, corpuscular variables $\lambda _0$
have a definite, but hidden, value (e.g., left or right slit) while
wave degrees of freedom, the amplitudes, have ge\-ne\-ri\-ca\-lly all possible components, although
they can interfere destructively in some marginals. In next section a particularly
relevant case, pairs of spin $1/2$ entangled particles with total null spin,
is analysed. 

The double role of amplitudes, as source of wave like effects and in Born's rule for 
observable relative frequencies, is behind the quantumness of QM in the proposed formulation, instead of a non local wave function collapse. There is no essential
distinction between entanglement and the two slit experiment.


\section{Applications}

\subsection{Two slit experiment}

The relevant variables in the two slit experiment are the slit variable $S$, with value
$L$ or $R$, and the position at the final screen $\bf R$, with values ${\bf r}_i$.
Slit and final position do not commute. Paths can be grouped in families
\footnote{Magnitudes associated to a path
are not necessarily values at its final point. Virtual paths going through $L$ and $R$ slits
are disjoint  families.} 
$[path](L,{\bf r}_i)$ and $[path](R,{\bf r}_i)$.
The extended Hilbert space is generated by elementary states $\|L,{\bf r}_i>$
and $\|R,{\bf r}_i>$. The vector of state is

\begin{equation}
\|S> = \sum _{i} Z(L, {\bf r}_i) \|L,{\bf r}_i>  + 
 \sum _{i} Z(R, {\bf r}_i) \|R,{\bf r}_i>  
\end{equation}
with projections 

\begin{equation}
\tau _S (\|S>) = \frac {1}{\sqrt{2}} (\|L> + \|R>)
\end{equation}
and

\begin{equation}
\tau _{\bf R} (\|S>) = \sum _i (Z(L, {\bf r}_i) + Z(R, {\bf r}_i))\|{\bf r}_i>
\end{equation}
This  one is also the orthodox vector of state in the 
position coordinates  representation at the final screen.

The formal distribution of probability 

\begin{equation}
P(L,{\bf r}_i)=|Z(L,{\bf r}_i)|^2 \quad
P(R,{\bf r}_i)=|Z(R,{\bf r}_i)|^2
\end{equation}
is not observable. The marginal

\begin{equation}
Z({\bf r}_i)=Z(L, {\bf r}_i) + Z(R, {\bf r}_i)
\end{equation}
determines the observable diffraction 
pattern $P({\bf r}_i)$ $=$ $|Z({\bf r}_i)|^2$.

Let us suppose that an external system interacts with our system of interest (particle plus de Broglie wave) at the $R$ slit. For simplicity,
let us consider that the new amplitude $Z'(R, {\bf r}_i)$ is similar in modulus to 
the unperturbed $Z(R, {\bf r}_i)$. However, there will be some phase shift
$Z'(R, {\bf r}_i, \phi) \simeq e^{i\phi} Z(R, {\bf r}_i)$. If we use, for example, a beam of coherent photons, and the additional system is  a phase plate of known phase shift, we will obtain a
displaced diffraction pattern. 

If the additional system is a (more or less complex) measurement apparatus, the phase shift will be unknown. A statistical average on the phase (e.g., with uniform distribution)
determines a total distribution of relative frequencies

\begin{equation}
P'({\bf r}_i) = \frac {1}{2\pi}\int _{-\pi}^{\pi}d\phi
|Z'(R, {\bf r}_i, \phi) +  Z(L, {\bf r}_i)|^2 \simeq
|Z(R, {\bf r}_i)|^2 + |Z(L, {\bf r}_i)|^2
\end{equation}
and the diffraction pattern disappears. We can observe now,
by correlation with the positive/negative result of measurement in slit $R$,
the marginal distributions $P'(R,{\bf r}_i)$ and $P(L,{\bf r}_i)$.
In practise, we can apply the projection rule according to the result of measurement 
at $R$ slit. 
A stochastic phase shift at measurement interactions  determines
the projection of state as a practical 
rule.

\subsection{Spin variables}

Let us consider a spin $1/2$ particle. The most general family of spin variables
will contain spin (up or down) in an arbitrary number of  directions
${\bf n}_j$, unit vectors in space. ${\cal F} = \{ S_1, S_2, \ldots \}$, with
$M_{S_j} = \{ +, - \}$ and $M_{\cal F} = \{ (s_1, s_2, \ldots ) \}$; $s_j$ is $+$ for spin up and $-$ for spin down in direction ${\bf n}_j$ \footnote{Direction $-{\bf n}_j$ is redundant, because $S_{-{\bf n}_j} $ $= - S_{{\bf n}_j} $.}.

The extended Hilbert space ${\cal H}_{ext}$ is generated by
elementary states $\|\lambda >$, $\lambda = (s_1, s_2, \ldots ) \in M_{\cal F}$. Orthodox Hilbert space is a two dimensional one ${\cal H}_{QM}$, where we can use each $\{|+_j>,|-_j>\}$ basis of spin up/down states  in  direction ${\bf n}_j$.

Let us associate a fix ``path'' amplitude to each spin value, $s_j e^{i\theta _j}$ for coplanar directions
or $s_j N_j$ for directions in space, with $N_j$ a quaternion number (with null real part) $N_j = n_j^x I + n_j^y J + n_j^z K$ associated to the 
vector ${\bf n}_j = n_j^x {\bf i} + n_j^y {\bf j} + n_j^z {\bf k}$ \cite{mio}.
The amplitude associated to an elementary state $\|\lambda>$ will be

\begin{equation}
Z(\lambda ) = \sum _j s_j N_j
\end{equation}

A states with known spin $s_1$ in direction ${\bf n}_1$ is

\begin{equation}
\|s_1> = \sum _{\pi _1(\lambda ) = s_1} Z(\lambda ) \|\lambda >
\end{equation}
Projection over the two dimensional Hilbert space for operator $S_1$, when using the basis $\{\|+_1>, \|-_1>\}$,
trivially reproduces the orthodox $|s_1>$: in the marginal amplitude, terms $s_j N_j$ 
and $-s_j N_j$ for $j>1$ cancel out, and a global  factor can be ignored.  

Projection over the Hilbert space for operator $S_2$, using the basis  $\{\|+_2>, \|-_2>\}$, gives

\begin{equation}
\tau _2 (\|s_1>) = {\cal N} \left( (s_1 N_1 + N_2) \|+_2> +
(s_1 N_1 - N_2) \|-_2> \right)
\end{equation}
with $\cal N$ a global factor. There is a correspondence, when appropriate 
phases are introduced in the correspondence of vectors, 
between $\tau _2 (\|s_1>)$ in the two dimensional quaternion Hilbert space
${\cal H}_{S_2} = < \|+_2>, \|-_2> >$ 
and the standard expression
in the orthodox complex Hilbert space.

All orthodox spin states of an individual particle are eigenstates of the spin operator
in some spatial direction. The former correspondence allows to
represent them in the extended phase space. On the other hand, the  label
$\lambda _0$ in the state $(\|s_1>, \lambda _0)$ will project onto a label $s_{1}$ (known for state $\|s_1>$)
or $s_{2}$ (unknown at the ensemble state $(s_1 N_1 + N_2) \|+_2> +$
$(s_1 N_1 - N_2) \|-_2>$). The (hidden) label determines the value of spin measurement
in arbitrary direction ${\bf n}_j$, $s_j = \pi _j(\lambda _0)$. 
 Born's rule in two steps for the extended formalism reproduces the standard results.
Operators $S_j$ commute. For dynamical purposes, additional operators 
$S_j^d$ fulfilling the standard $[S_x^d, S_y^d]$ $= i \hbar S_z^d$ should be considered.

Let us consider a composite system of two spin $1/2$ particles 
${\cal S}^a$ and ${\cal S}^b$ in a total null spin state.
The label $\lambda _0$ will fulfil $\pi _j(\lambda _{a0})$ $+$ $\pi _j(\lambda _{b0}) = 0$,
i.e., perfect correlation $s_j^a + s_j^b = 0$, determining a subset $M_{corr} \subset M_{\cal F}$;
the total family of operators is ${\cal F} = {\cal F}^a \cup {\cal F}^b$,
${\cal F}^a = \{ S^a_1, S^a_2, \ldots \}$ and ${\cal F}^b = \{ S^b_1, S^b_2, \ldots \}$.

The vector of state for the composite will have the form

\begin{equation}
\|S> = \sum _{\lambda \in M_{\cal F}} Z_T(\lambda ) \|\lambda ^a>\|\lambda ^b>
\end{equation}
Notice that $\lambda$ is not restricted to $M_{corr}$, all we have to 
impose is that marginals $Z(s_1^a,s_1^b=s_1^a) = 0$, through destructive interference.
A solution for the amplitude distribution is

\begin{equation}
Z_T(\lambda ) = Z_T(\lambda ^a, \lambda ^b ) =
Z(\lambda ^a) - Z(\lambda ^b)
\end{equation}
where $Z(\lambda )$ is as before $Z(\lambda ) = \sum _j s_j N_j$. 

Let us consider two directions ${\bf n}_1$ and ${\bf n}_2$, with variables $(s^a_1, s^a_2, s^b_1, s^b_2)$.
The marginal amplitude becomes

\begin{equation}
Z_{(1,2)^a,(1,2)^b}(s^a_1, s^a_2, s^b_1, s^b_2) = 
(s^a_1-s^b_1) N_1 + (s^a_2-s^b_2)N_2
\end{equation}
up to a total factor; all coefficients of $N_j$ for $j>2$ vanish.
Therefore, $Z_{(1,2)^a,(1,2)^b}$ $(s^a_1, s^a_2, s^b_1 = s^a_1, s^b_2)$ $=$ 
$(s^a_2-s^b_2)N_2$ $\neq 0$. However,  the observable amplitudes
of measurement in direction ${\bf n}_1$ for both particles  is

\begin{equation}
Z_{1^a,1^b}(s^a_1,s^b_1) = 
(s^a_1-s^b_1) N_1 
\end{equation}
We get $Z_{1^a,1^b}(s^a_1,s^b_1= s^a_1) = 0$, $Z_{1^a,1^b}(s^a_1,s^b_1= -s^a_1)$ 
$= 2s^a_1 N_1$,
and a probability distribution
$P(s_1,s_1) = 0$, $P(s_1,-s_1) = 1/2$.

Similarly, the observable amplitude $Z_{1^a,2^b}$ for measurement $S_1$ of particle $a$ and $S_2$ of particle $b$ is

\begin{equation}
Z_{1^a,2^b}(s^a_1,s^b_2) = s^a_1 N_1 - s^b_2 N_2
\end{equation}
with the quantum associated probability distribution
$P(s^a_1,s^b_2) ::$ $|s^a_1 N_1 - s^b_2 N_2|^2$ $=$ $2(1 - s^a_1 s^b_2\,\, {\bf n}_1 \cdot {\bf n}_2)$ 
\footnote{$N_1^* = - N_1$ 
for quaternions without real part, and $N_1^* N_2 =$
$- {\bf n}_1 \cdot {\bf n}_2$ $ - {\bf n}_1 \times {\bf n}_2$; 
in the former expression, the dot product is the real part and the cross product
the imaginary part of a quaternion.}.

If we calculate the marginal amplitude for direction $S_1$ of particle $a$ and
directions $S_2$ and $S_3$ of particle $b$

\begin{equation}
Z_{1^a,(2,3)^b}(s^a_1,(s_2,s_3)^b) = s^a_1 N_1 - (s^b_2 N_2 + s^b_3 N_3)
\end{equation}
we can define a formal distribution of probability 

\begin{equation}
P(s^a_1,(s_2,s_3)^b) =
|Z_{1^a,(2,3)^b}(s^a_1,(s_2,s_3)^b)|^2
\end{equation} 
The marginal probability
$P'(s_1,s_2)$  $=$ $P(s^a_1,(s_2,+_3)^b)$ $+$ $P(s^a_1,(s_2,-_3)^b)$ does not match the former one, because of interference
when applying Born's rule in two steps,

\begin{equation}
Z_{1^a,2^b}(s^a_1,s^b_2) = Z_{1^a,(2,3)^b}(s^a_1,(s_2,+_3)^b) + 
Z_{1^a,(2,3)^b}(s^a_1,(s_2,-_3)^b)
\end{equation}
and 

\begin{equation}
P(s^a_1,s^b_2) :: |Z_{1^a,(2,3)^b}(s^a_1,(s_2,+_3)^b) + 
Z_{1^a,(2,3)^b}(s^a_1,(s_2,-_3)^b)|^2 
\end{equation}
differs, even projectively, from

\begin{equation}
|Z_{1^a,(2,3)^b}(s^a_1,(s_2,+_3)^b)|^2 + |Z_{1^a,(2,3)^b}(s^a_1,(s_2,-_3)^b)|^2
\end{equation}

There is a full analogy with interference in the two slit experiment,
for exam\-ple if we identify the third magnitude $s_3 \in \{+, -\}$ with the slit $\{L,R\}$. Bell's inequalities
prove that there is no distribution of probability in a space with hidden variables
whose marginal probabilities reproduce the quantum result. What we have  here is a distribution of amplitude in the extended space of hidden varia\-bles; through the
defined correspondence (marginals, interference, Born's rule) it reproduces the orthodox quantum phase space
representation. 

Corpuscular 
variables $\lambda _0$ fulfil the  perfect correlation condition, and determine the result of arbitrary measurements in a correlated pair of particles
$(\|S>,$ $(\lambda _{a0},$ $\lambda _{b0}))$. The distribution 
of probability for observable  magnitudes depends on the wave degrees of freedom $\|S>$
(common for an ensemble), the distribution of 
amplitudes. Born's rule in two steps gives way to interference, wave like phenomenon that does not appear in a marginal probability.
The distribution $Z(\lambda)$ of the entangled system, as well as the corpuscular 
variable $\lambda _0$, are fixed from the generation event of the two entangled particles.
No non local phenomenon is invoked 
in the former mathematical description.

\section{Summary and Outlook}

A formalism of non relativistic Quantum Mechanics in which elementary states
have joint precise values of non commuting magnitudes has been developed in order to
restore the action reaction principle, which is
contradictory with the projection of state. 

An abstract path integral formalism, with families of virtual paths more res\-tric\-tive than the
standard one, suggest that a correspondence with all predictions of the orthodox theory could 
exist. The two slit experiment and the phase space of spin variables have been formulated in the extended phase space.

Contextuality and non locality of Quantum Mechanics appear here
as interference properties of  wave like degrees of freedom. The mysterious
double role of amplitudes, at interference and determining probabilities,
is the source of these non classical properties of quantum systems.

An accompanying, wave like subsystem is predicted 
when particle and detector are spatially separated in indirect measurements, in order to  preserve
locality of interactions and the action reaction principle. An appropriately designed detector
could show observable reactions; they would be assigned  either to a non local interaction 
between the corpuscular component and the detector or to
a local interaction between a de Broglie wave component and detector.


\section{Acknowledgements}

Financial support from research project MAT2011-22719 is acknow\-ledged. I also
kindly ack\-now\-led\-ge helpful comments from members of the audience in both
seminars at Zaragoza and Valladolid Universities, where I presented some
conclusions of this research in April and June 2015 respectively.


\end{document}